# Kinetic Suppression of Photoinduced Halide Migration in Wide Bandgap Perovskites via Surface Passivation

*Farhad Akrami[1], Fangyuan Jiang[1], Rajiv Giridharagopal[1], David S. Ginger[1,2*]*

[1]Department of Chemistry, University of Washington, Seattle, WA 98195, USA

[2]Physical Sciences Division, Physical and Computational Sciences Directorate, Pacific Northwest National Laboratory, Richland, WA 99352, USA

\* Corresponding author: dginger@uw.edu

In this work, we study the kinetics of photoinduced halide migration in $FA_{0.8}Cs_{0.2}Pb(I_{0.8}Br_{0.2})_3$ wide (~1.69 eV) bandgap perovskites and show halide migration slows down following surface passivation with (3-aminopropyl) trimethoxysilane (APTMS). We use scanning Kelvin probe microscopy (SKPM) to probe the contact potential difference (CPD) shift under illumination, and the kinetics of surface potential relaxation in the dark. Our results show APTMS-passivated perovskites exhibit a smaller CPD shift under illumination, and a slower surface potential relaxation in the dark. We compare the evolution of the photoluminescence spectra of APTMS-passivated and unpassivated perovskites under illumination. We find that APTMS-passivated perovskites exhibit more than 5 times slower photoluminescence redshift, consistent with the slower surface potential relaxation as observed by SKPM. These observations provide evidence for kinetic suppression of photoinduced halide migration in APTMS-passivated samples, likely due to reduced halide vacancy densities, opening avenues to more efficient and stable devices.

**TOC Graphic**

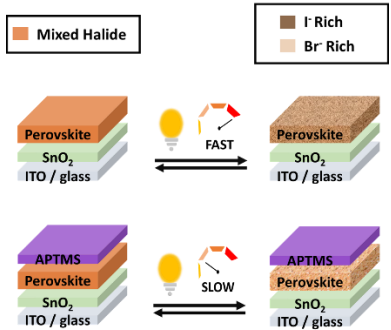

Metal halide perovskites are being widely explored in applications ranging from solar cells and light-emitting diodes to sources of quantum light.[1–5] The research community is demonstrating rapid advancements in the performance of these devices. For instance, in 2023, the perovskite solar cell power conversion efficiency (PCE) record stands at 26.1% for single junction and 33.7% for perovskite-silicon tandems, compared to 23.3% and 27.3% only 5 years ago.[6] At the same time, improved materials have enabled efficient perovskite light-emitting diodes,[3] and even coherent single photon sources.[5]

While perovskites are attractive semiconductor materials because they are easily synthesized and processed over large areas at low temperatures, these properties make them prone to a range of surface and bulk defects.[7–11] Understanding and controlling these defects, particularly unpassivated surface states, is important due to their influence on nonradiative loss pathways that affect device performance and stability.[12–16] Furthermore, certain defects like halide vacancies also facilitate ionic conductivity, for instance via vacancy migration.[17–21] Passivation strategies that remove halide vacancies could thus have important implications for materials performance and stability, especially for the mixed halide perovskites that are commonly used to tailor the semiconductor bandgap in solar cell and light-emitting diode applications.[2,3,22,23]

Many groups have studied surface passivation strategies, usually with an emphasis on reducing non-radiative recombination as probed by photoluminescence.[11,24–27] Others have focused on photoinduced ion migration in both pure halide and mixed halide compositions.[28–33] However, despite the need for understanding and controlling the kinetics of photoinduced ion migration, experimental evidence for kinetic effects of surface passivation on photoinduced halide migration is limited.[34–36]

Previously, our group has shown that (3-aminopropyl) trimethoxysilane (APTMS) can successfully reduce nonradiative recombination,[11] improve the open-circuit voltage of devices,[37] and mitigate electric field-induced ion motion in perovskite films.[38] Based on the role of halide vacancies in ion migration,[17–21] we hypothesize that surface passivation should also kinetically hinder *photoinduced* ion migration by reducing the density of halide vacancies. In this work, we test this hypothesis by studying the kinetic effects of APTMS surface passivation on photoinduced halide migration in $FA_{0.8}Cs_{0.2}Pb(I_{0.8}Br_{0.2})_3$, a wide bandgap perovskite ($E_g$ of 1.69 eV). We chose this formulation as it offers an ideal optical bandgap for top sub-cells in perovskite-silicon tandems. We study the effects of passivating the perovskite surfaces with APTMS by tracking both the contact potential difference (CPD) shift and photoluminescence evolution with time under illumination, and the kinetics of surface potential relaxation in the dark following illumination. Consistent with our hypothesis, we observe that APTMS-treated samples undergo a smaller CPD shift under illumination and exhibit a slower surface potential relaxation in the dark. In addition, the APTMS-treated samples show more than 5 times slower redshift in photoluminescence under illumination.

We synthesize $FA_{0.8}Cs_{0.2}Pb(I_{0.8}Br_{0.2})_3$ perovskites by adapting previously reported methods[39] as described in detail in the Supporting Information (SI). To passivate the surface of these samples, we deposit APTMS using a vacuum oven for 5 minutes, as in our previous work (see SI).[11] We measure the bandgap to be ~1.69 eV via UV-Vis spectroscopy (Figure S1a), and confirm the perovskite structure via X-ray diffraction (XRD) (Figure S1b). Figure S1b shows that there are no detectable changes to the perovskite XRD patterns after APTMS surface passivation. We confirm the successful surface passivation of this perovskite composition by APTMS via steady-state and time-resolved photoluminescence, which show significant increases in photoluminescence intensity (~36 times increase) and photoluminescence lifetime (~7 times increase) (Figure S3 – 4).

To further study the effects of perovskite surface passivation with APTMS, we use scanning Kelvin probe microscopy (SKPM) to probe the CPD shift under illumination and the kinetics of surface potential relaxation in the dark following illumination (Figure 1). Figure 1a shows a schematic diagram of the samples used for the SKPM and steady-state photoluminescence measurements in this study (see SI for details). Briefly, we deposit the perovskite films on spin coated $SnO_2$ electron transport layer on top of indium tin oxide (ITO) covered glass substrates. Figure 1a also shows the molecular structure of the APTMS monomer, and the structure of the first layer of polymerized APTMS as it is likely to interact with the perovskite.[11] Figure 1b depicts a simplified SKPM measurement setup employed in this study.

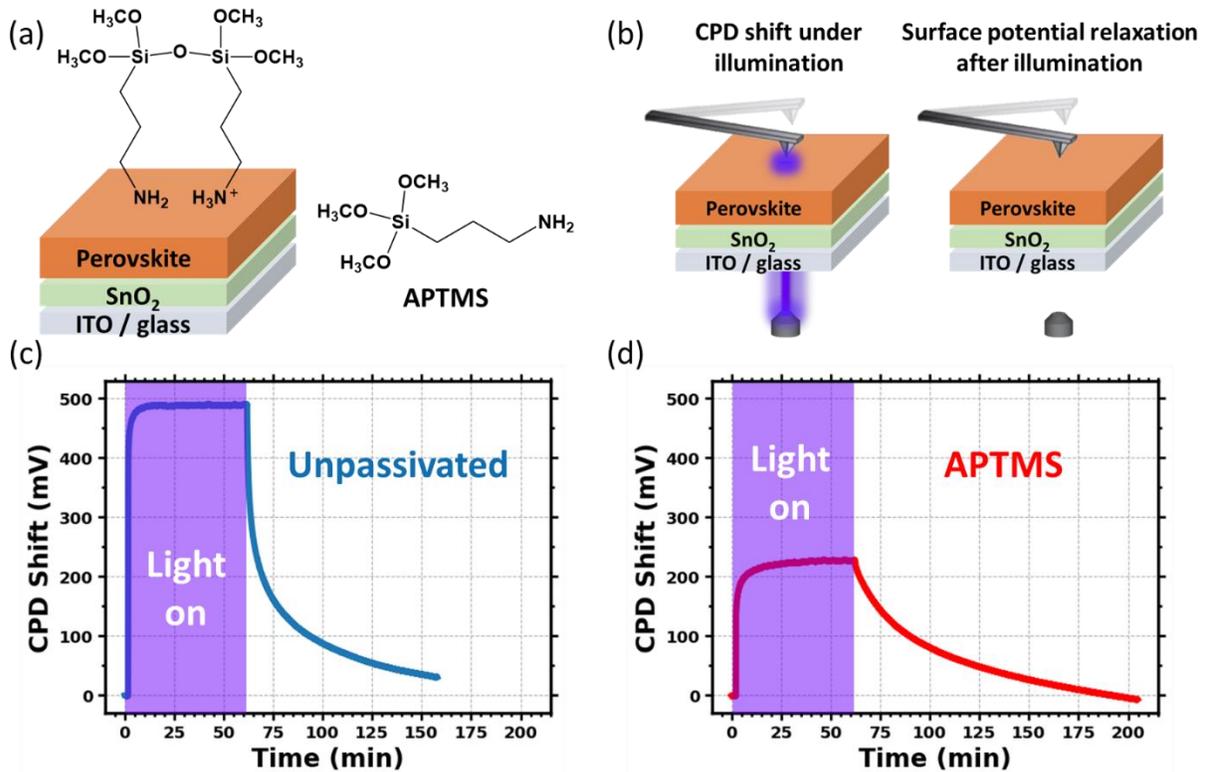

Figure 1. SKPM of unpassivated and APTMS-passivated samples. (a) A schematic diagram of samples and the molecular structure of APTMS surface passivator shown as a monomer (on the right) and polymerized as it is likely to interact with the perovskite (on the top). (b) A simplified SKPM measurement setup on the surface of a sample. CPD shifts as a function of time showing the effects of illumination on CPD evolution and surface potential relaxation following illumination of (c) unpassivated, and (d) APTMS-passivated samples. The illumination source is a 405 nm continuous wave laser, 18 kW m$^{-2}$, co-aligned with the atomic force microscopy (AFM) tip.

Figure 1c – d compare the CPD shift in unpassivated and APTMS-passivated samples as a function of time during and after illumination. As shown in Figure 1c – d, APTMS surface passivation leads to a smaller CPD shift under illumination, and slower surface potential relaxation after switching off the illumination source. We use a stretched exponential decay function to determine the surface potential relaxation time in unpassivated perovskite to be ~1000 s (~16.7 min) while in APTMS-passivated sample it is ~2550 s (~42.5 min) (see SI and Table S1 for details). We also observe a smaller CPD shift under illumination and slower surface potential relaxation in APTMS-passivated samples at a higher illumination intensity (Figure S5 – 6). Notably, we observe a fast early surface potential relaxation in unpassivated samples at both light intensities which is less pronounced in APTMS-passivated samples.

We propose that the differences in CPD shift magnitude under illumination for the unpassivated and APTMS-treated samples is primarily because of differences in surface charge accumulation, whereas ion motion explains the time-dependence during and after illumination. Prior to illumination, the higher density of surface defects causes a larger surface band bending in unpassivated samples. Upon illumination, the surface defects cause a larger surface charge accumulation in unpassivated samples which manifest in the higher CPD shifts observed. APTMS surface passivation decreases the density of surface trap states and hence less surface charge accumulation is expected.

One of the major challenges in studying ion motion in halide perovskites is coupled ionic and electronic motion, though studies suggest these processes can be decoupled due to differences in their timescales.[40–42] Electronic motion is reported to occur with a timescale of femtoseconds to microseconds, while ion motion is reported to occur with milliseconds to seconds and longer timescales.[40–42] Surface potential relaxation occurs on a timescale of minutes which is consistent with ion motion. Furthermore, due to the lower activation energy of mobile halides[17,20] and our previous visualization of photo-induced halide migration in hybrid organic-inorganic perovskites,[21] we expect that the dominant mobile ions are halides. Therefore, we attribute surface potential relaxation to halide migration. After illumination, the light source is switched off which allows for halide ions to slowly migrate back towards an equilibrium state. This process is thermodynamically driven by entropy, and kinetically governed by concentration gradients.[14,21,43] Due to higher surface charge accumulation in unpassivated samples, initially the trapped halides migrate back relatively quickly. However, new lower concentration gradients slow down the kinetics over time, as shown in Figure 1c – d and Figure S6. Less surface charge accumulation in APTMS-passivated samples mitigate the early fast surface potential relaxation observed in unpassivated samples and suppress the overall kinetics (see SI for details).

Based on the slower kinetics of APTMS-passivated samples as probed by SKPM, we expect that APTMS-treated samples should also exhibit kinetically suppressed halide phase segregation. To test this, we conduct photoluminescence measurement because hole funneling into lower-bandgap iodide rich domains makes halide phase segregation easily detectable via this method.[14,29] We monitor the photoluminescence $\lambda_{max}$ shift under illumination as a function of time (Figure 2). Figure 2 shows that APTMS passivation helps suppress the photoluminescence redshift that occurs under illumination in these samples. We also study the intensity dependence of this process. As shown in Figure 2c, higher illumination intensities cause faster photoluminescence redshifts in both unpassivated and APTMS-passivated samples. However, APTMS-passivated samples exhibit more than 5 times slower photoluminescence redshift. Notably, APTMS-passivated samples undergo slower photoluminescence redshift even at the illumination intensity of 46 kW

m$^{-2}$ compared to unpassivated samples at both 23 kW m$^{-2}$ and 46 kW m$^{-2}$. This observation shows the effectiveness of APTMS passivation in mitigating halide segregation.

Although we cannot completely rule out the possibility of deeply trapped electronic carrier's motion, both SKPM and photoluminescence data show kinetics with timescales on the order of minutes, and the PL shifts are a clear signature of ion migration. Considering the mixed A-site cations in our studied perovskite composition and the possibility of A-site cation phase segregation as reported by different groups,[44–46] we also make pure iodide films with the same ratio of the mixed A-site cations, FA$_{0.8}$Cs$_{0.2}$PbI$_3$, and probe photoluminescence of this composition under illumination (Figure S7 – 8). We observed no significant photoluminescence $\lambda_{max}$ shift, further suggesting that the observed photoluminescence peak shift in the mixed halide samples is due to halide phase separation.

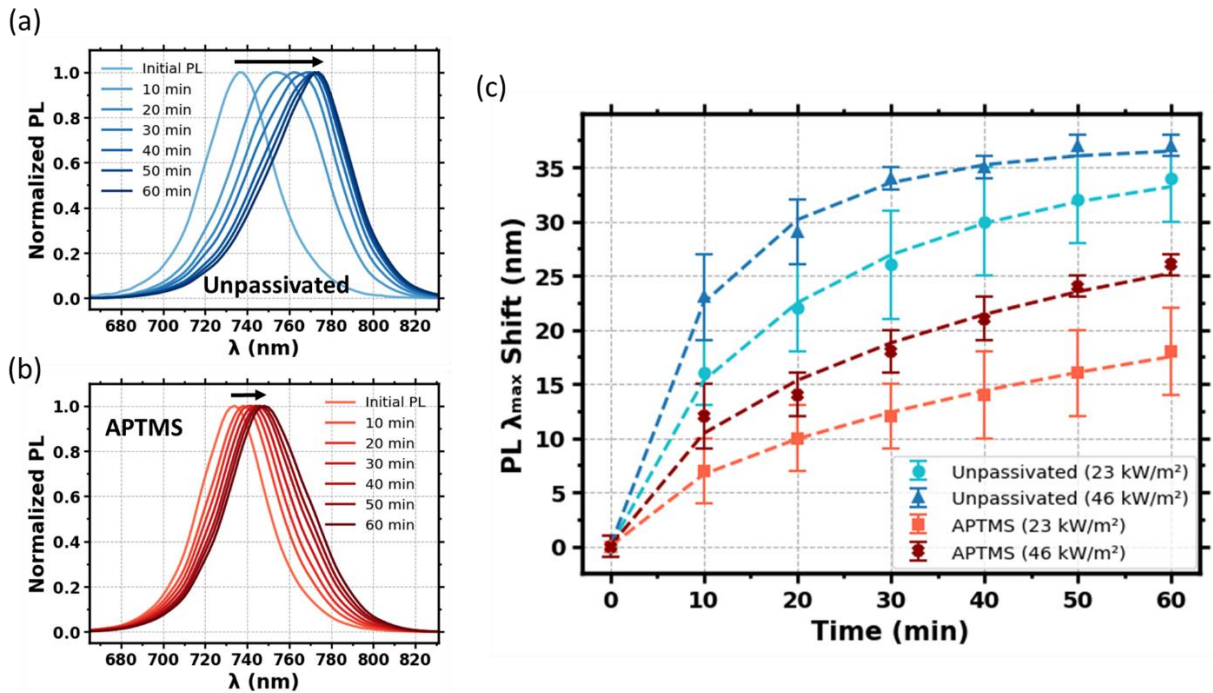

Figure 2. Photoluminescence of unpassivated and APTMS-passivated samples under illumination. Photoluminescence as a function of light soaking time with intensity of 23 kW m$^{-2}$ of (a) unpassivated and (b) APTMS-passivated perovskites. (c) Photoluminescence $\lambda_{max}$ shift as a function of light soaking time of unpassivated with intensity of 23 kW m$^{-2}$ (light blue), 46 kW m$^{-2}$ (dark blue), and APTMS-passivated perovskites at 23 kW m$^{-2}$ (light red) and 46 kW m$^{-2}$ (dark red). The light source used is a 532 nm continuous wave laser. The dashed lines are the fits for each curve (see SI for details). Error bars are standard error of the mean for three measurements performed on three different films.

As shown in Figure 2a, continuous light soaking under constant intensity exacerbates the halide segregation over time, showing the impact of light soaking time. Similarly, in Figure 3a, we examine the effect of light intensity under constant soaking time. We observe that higher intensities cause more halide segregation. Thus, both light soaking time and intensity are factors that contribute to halide segregation. To compare the impact of these two factors, we plot photoluminescence $\lambda_{max}$ shift as a function of light fluence (kJ cm$^{-2}$). Figure 3b shows that within both unpassivated and APTMS-passivated samples, continuous light soaking with the same fluence (kJ cm$^{-2}$) regardless of time/light intensity (23 or 46 kW m$^{-2}$) causes similar amount of photoluminescence redshift. We also probe the effect of light soaking on light absorption of the perovskite sample and find that light soaking does not affect the light absorption of the sample (Figure S9), likely due to small fraction of the mixed halide phase undergoing halide phase segregation which may not be as easily detectable via absorption measurements.[47]

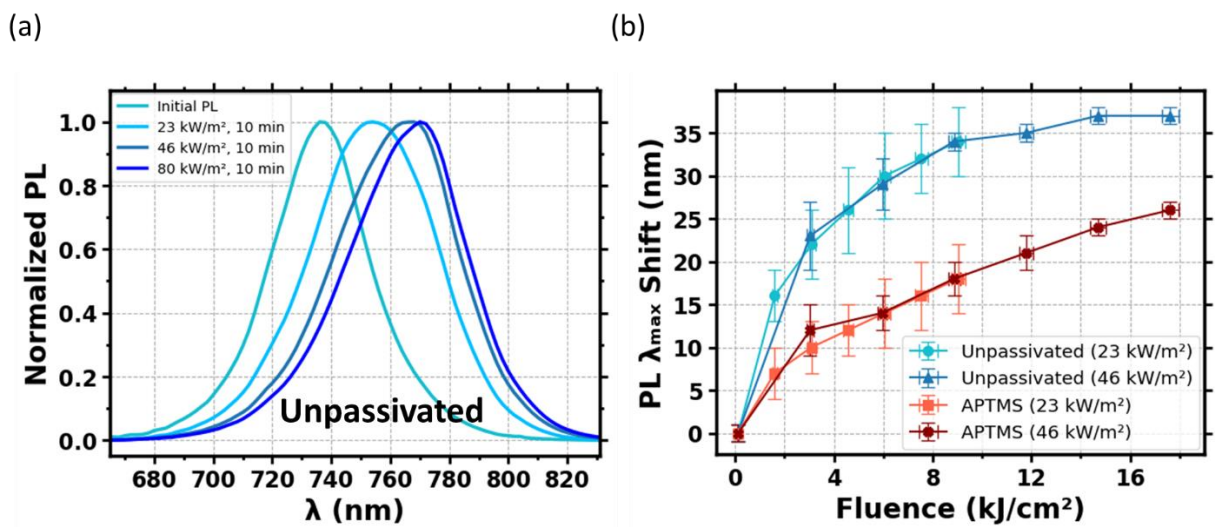

Figure 3. Effect of light soaking time and intensity on halide segregation. (a) Photoluminescence of unpassivated perovskites as a function of light intensity with constant light soaking time of 10 min. (b) Photoluminescence $\lambda_{max}$ shift as a function of light fluence (kJ cm$^{-2}$) of unpassivated with light intensity of 23 kW m$^{-2}$ (light blue), 46 kW m$^{-2}$ (dark blue), and APTMS-passivated perovskites at 23 kW m$^{-2}$ (light red) and 46 kW m$^{-2}$ (dark red).

Our findings in Figure 3 suggest that photoinduced halide segregation follows first order kinetics, with halide segregation dependent on light fluence (kJ/cm$^2$), expressed in reaction 1 below (as a limiting case).

$$5FA_{0.8}Cs_{0.2}Pb(I_{0.8}Br_{0.2})_3 + h\nu \rightleftharpoons 4FA_{0.8}Cs_{0.2}PbI_3 + FA_{0.8}Cs_{0.2}PbBr_3 \quad (1)$$

This shows that when examining photoinduced halide segregation, special attention should be given to the light fluence (the product of light intensity and time), as both light intensity and time contribute to photoinduced halide segregation.

In summary, we report that APTMS, applied as a surface passivator, kinetically suppresses halide migration in $FA_{0.8}Cs_{0.2}Pb(I_{0.8}Br_{0.2})_3$, a wide bandgap perovskite ($E_g$ of 1.69 eV). We conduct SKPM to show that APTMS surface passivation kinetically suppresses halide relaxation in the dark after illumination. Using photoluminescence, we investigate the kinetic effects of APTMS surface passivation on halide phase segregation and show that APTMS-passivated samples exhibit more than 5 times slower halide segregation. Lastly, we report that halide segregation follows first order kinetics dependent on light fluence (kJ/cm$^2$), where the same light fluence leads to the same extent of halide segregation regardless of time/light intensity. This study shows that passivating surface defects has more implications beyond reducing electronic carrier recombination rates, and provides evidence for kinetic suppression of photoinduced halide migration which is a crucial topic for improving efficiency and stability of perovskites.

**Experimental Methods**

All information regarding the materials, sample preparations, and characterizations (AFM topography, UV-Vis, XRD, and photoluminescence of perovskites etc.) can be found in Supporting Information.

**Notes**

The authors declare no competing interests.

**Acknowledgments**

This paper is based primarily on work supported by the U.S. Department of Energy (DOE-SC0013957). The authors acknowledge the use of facilities and instruments at the Photonics Research Center (PRC) at the Department of Chemistry, University of Washington, as well as that at the Research Training Testbed (RTT), part of the Washington Clean Energy Testbeds system. Part of this work is conducted at the Molecular Analysis Facility, a National Nanotechnology Coordinated Infrastructure site at the University of Washington which is supported in part by funds from the National Science Foundation (awards NNCI-2025489, NNCI-1542101), the Molecular Engineering & Sciences Institute, and the Clean Energy Institute. We acknowledge Dr. Tanka R. Rana and Prof. J. Devin MacKenzie from the University of Washington for providing the commercial SnO$_2$ solution. D.S.G. acknowledges salary and infrastructure support from the Washington Research Foundation, the Alvin L. and Verla R. Kwiram Endowment, and the B. Seymour Rabinovitch Endowment.

**Supporting Information Available:** The supporting information is available at xxx (PDF).